\documentclass[prl,twocolumn,showpacs]{revtex4}

\usepackage{graphicx}%
\usepackage{dcolumn,float}
\usepackage{amsmath,amssymb}

\begin{document}

\title{Renormalization Group study of the Sliding Luttinger liquids}

\date{\today}

\author{  Germ\'an  Sierra} 

\affiliation{ 
Instituto de Matem\'aticas y F\'{\i}sica Fundamental, C.S.I.C.,
Madrid, Spain. }

\begin{abstract}
We derive the RG-flow equations of  
the sliding Luttinger liquid perturbed by 
charge-density-wave (CDW) and superconducting (SC) operators.
Using them we study the phase diagram of an array of XXZ
spin chains coupled by Ising terms. In the weak coupling regime 
we find a new class of non-gaussian and unstable fixed points 
whose existence is due to a balance
between the CDW, SC and sliding Luttinger couplings.  
\end{abstract}


\maketitle

\newcommand{\bb}{\boldsymbol{\beta}}
\newcommand{\ba}{\boldsymbol{\alpha}}


The Luttinger liquid (LL) is the key concept
to describe interacting electrons in one dimension. 
The collective nature of the excitations,
the spin-charge separation and the power law behaviour 
of correlators with anomalous exponents
are some of its distintive feautures, 
in contrast to the Fermi liquid
theory \cite{emery}. Much of the theoretical
activity in the last years has been devoted to
the search of non-Fermi liquid theories in dimensions
higher than one, specially in 2D, due to its possible
connection with the high-$T_c$ superconductors 
and another strongly correlated systems. 

A natural path  to realize a non-Fermi liquid
has been to couple arrays of 1D Luttinger liquids
forming ladders and  2D planes. 
However the general consensus has been, until recently, 
that 2D arrays of LL's are unstable to the formation of
crystal, superconducting  or 2D Fermi liquid
states \cite{schulz}. An alternative to these ``no-go theorems''
has been proposed lately using the concept
of sliding Luttinger liquid (SLL), 
also called smectic non-Fermi liquid 
\cite{emery00,carpentier,lubensky01}.

These works were partially motivated by the Anderson's
proposal of confinement of excitations  in the
Luttinger liquids \cite{anderson} and have a classical analogue
in the stacks of coupled 2D XY models coupled
by gradient interactions \cite{dna}.  The SLL model
may also be relevant to the stripe phases
of the Quantum Hall effect and the cuprates.

The sliding Luttinger liquid is the fixed 
point of a gaussian Hamiltonian which 
treats on equal footing the individual Luttinger Hamiltonians
of the stripes and the density-density 
and current-current interstripe interactions \cite{emery00,carpentier}. 
Using bosonization techniques one can regard
the SLL as a set of decoupled LL's
characterized by a sound velocity $v(q_\perp)$ and 
a Luttinger coupling $K(q_\perp)$ which depend
on the tranverse momentum $q_\perp$ across
the stripes. The gaussian nature of the SLL fixed
point allows a simple
derivation of the scaling dimensions of the 
single particle (SP), 
charge density (CDW) and superconducting (SC) 
order parameters, in terms of the SLL
function $K(q_\perp)$ \cite{emery00,carpentier}. These scaling 
dimensions have been used 
to study the stability of the SLL under
various perturbations, finding rich
phase diagrams where the SLL phase survives 
in the vicinity of CDW, SC and Fermi liquid
phases  \cite{emery00,carpentier,lubensky01}. 
The perturbative  RG analysis performed in 
the latter references take into account  the running 
of the coupling constants of the perturbations
to first order,  while the Luttinger functions $K(q_\perp)$
and $v(q_\perp)$ stay constant. It is however well
known that in certain situations, 
as in the presence of  marginal
perturbations, one has to consider in addition  
the renormalization  of the Luttinger parameters
in a RG \`a la Kosterlitz-Thouless (KT) \cite{KT}. 
This is for example the case of the Hubbard model
at half filling where the Umklapp operator
becomes marginally relevant for a repulsive 
Hubbard constant, leading to a charge gap in the
low energy spectrum. 

The aim of this Letter is to derive the 
RG equations of the Sliding Luttinger liquid
for the coupling constants and the SLL 
functions to second order in the couplings, 
and study some of their consequences in a
model consisting of  arrays 
of spinless 1D Hamiltonians coupled 
by density-density interactions. The latter
model has been treated in the past with 
bosonization \cite{jose1}, mean field \cite{jose2} 
and variational
methods \cite{tpa}, which predict the existence of large
regions in parameter space where the phase is 
either CDW-like  or XY-like, corresponding 
respectively to the smectic crystal and
the smectic metallic phases of references \cite{emery00,carpentier}.

On more general grounds 
we also analyse the
stability of the spinless SLL under  marginal  CDW and SC
perturbations,  reaching the conclusion
that when one of these perturbations is irrelevant
the other one is relevant and hence the SSL 
is an unstable fixed point. 
We also find new
unstable fixed points if two conditions are
satisfied:  i) when 
the CDW coupling constants are minus the superconducting
ones, which guarantees the freezing of  
the SLL parameters, and ii) when the running of the
CDW and SC parameters to first order, which is given
by their scaling dimension,  is cancelled out by
the second order term. The mechanism involved here 
is similar to the one giving rise to the  Wilson-Fisher
fixed point in dimensions $d < 4$ \cite{cardy}.
From a more mathematical viewpoint the manifold of 
these non gaussian
fixed points are closely related to the target manifolds
that appear in 5 dimensional 
supergravity theories \cite{5d}.

Let us consider an array of $N$ spinless Luttinger stripes
with phase fields for the 
density fluctuations $\phi_a\; ( a= 1, \dots, N)$ and 
euclidean Lagrangian 

\begin{equation}
{\cal L}_0 = \sum_{a=1}^N \frac{K_0}{2} \left[ \frac{1}{v_0} 
(\partial_t \phi_a)^2 + v_0 (\partial_x \phi_a)^2 \right]
\label{1} 
\end{equation}

\noindent where $K_0$ is the inverse of the standard Luttinger
parameter ( $K_0 > 1 $ for repulsion) and $v_0$ is the sound
velocity which we scale to 1. 
The charge density fluctuations $j_0^a$ and the charge currents 
$j_1^a$ are given by the bosonization equation $j_\mu^a = \frac{1}{\pi}
\epsilon_{\mu \nu} \partial^\nu \phi_a$. Hence the density-density
and current-current interactions among the stripes are also
quadratic in the derivatives of the bosonic fields and, together
with (\ref{1}),  define the  sliding Luttinger 
Lagrangian \cite{emery00,carpentier}

\begin{align}
{\cal L}_{SLL} = &  \frac{1}{2}  \sum_{a,b=1}^N \left[  
\partial_t \phi_a K^J_{a,b} \partial_t \phi_b
+ \partial_x \phi_a K^\rho_{a,b}  \partial_x \phi_b \right]
\label{2} 
\end{align}

\noindent where $K^{J,\rho}_{a,b} = \delta_{a,b} K_0 + \bar{K}^{J,\rho}_{a,b}$
are $N \times N$ matrices
whose off-diagonal elements are given by the interstripe 
current-current and density-density interactions. 
The SLL model can alternatively be formulated in the dual variables
$\theta_a$,  which are the phase fields of the superconducting
fluctuations. The SLL Lagrangian (\ref{2}) becomes 
\cite{emery00,carpentier,lubensky01}

\begin{align}
{\cal L}_{SLL} = &  \frac{1}{2}  \sum_{a,b=1}^N \left[  
\partial_t \theta_a (K^{-1}_\rho)_{a,b} \partial_t \theta_b
+ \partial_x \theta_a (K_J^{-1})_{a,b}  \partial_x \theta_b \right]
\label{4} 
\end{align}

\noindent where $K^{-1}_{J,\rho}$ are the inverse matrices
of $K^{J,\rho}$. Both eqs.(\ref{2}) and (\ref{4}) exhibit 
the smectic or sliding 
symmetries $\phi_a \rightarrow
\phi_a + \alpha_a$ and $\theta_a \rightarrow \theta_a + \beta_a$, 
where $\alpha_a$ and $\beta_a$ are constants, which prevents
the lock in of the charge-density-wave and superconducting
order parameters of the individual stripes \cite{emery00,carpentier}. 
Assuming periodic boundary conditions across the stripes
and translational invariance along them, one can perform 
the Fourier transform:  
 $\phi_a = \frac{1}{\sqrt{N}} \sum_{q_\perp} e^{i q_\perp a} \phi_{q_\perp} , \;\;
K^{J,\rho}(q_\perp) = \sum_{q_\perp} e^{i q_\perp a} K^{J,\rho}_{1,1+a}$
in order to bring the SLL Lagrangian (\ref{2}) into the form

\begin{equation}
{\cal L}_{SLL} = \frac{1}{2} \sum_{q_\perp} K(q_\perp) 
\left[ \frac{1}{v(q_\perp)} |\partial_t \phi_{q_\perp} |^2 
+ {v(q_\perp)} |\partial_x \phi_{q_\perp} |^2 \right]
\label{6}
\end{equation}

\noindent where $K(q_\perp) = \sqrt{ K^J(q_\perp) K^\rho(q_\perp) }$
is the (inverse) Luttinger parameter 
and   $v(q_\perp) = \sqrt{ K^\rho(q_\perp) /K^J(q_\perp) }$
is the sound  velocity  
of the $q_\perp-$mode. In the dual variables $\theta_{q_\perp}$
the Lagrangian has the same form as eq.(\ref{6}) 
with the replacement $K(q_\perp) \rightarrow 1/K(q_\perp)$. 
 The scaling dimensions of the various 
operators, given in references \cite{emery00,carpentier}, 
can be easily
derived by observing that a  generic  vertex operator 
$V_{\bb}^\phi = {\rm exp}(i \sum_a \beta_a \phi_a)$, 
factorizes into the product of vertex
operators on  the $q_\perp-$modes as follows

\begin{align}
V_{\bb}^\phi&  = \prod_{0 \leq q_\perp \leq \pi} 
e^{ i \sqrt{\frac{2}{N}} \sum_a \beta_a ( {\rm cos} (q_\perp a) 
\phi_{q_\perp}^{(1)} + {\rm sin} (q_\perp a) 
\phi_{q_\perp}^{(2)} ) } \nonumber  \\
\phi_{q_\perp}& = \frac{1}{\sqrt{2}}  (\phi_{q_\perp}^{(1)} 
- i \phi_{q_\perp}^{(2)}), 
\;\; ( q_\perp \neq 0 , \pi) \label{7} 
\end{align}

\noindent where $\phi_{q_\perp}^{(1,2)}$ are real scalar fields.
The scaling dimension $\Delta_{\bb}^\phi$  of $V_{\bb}^\phi$   
is given simply by the sum 
of  the scaling dimensions of all the individual $q_\perp-$modes, namely

\begin{align}
\Delta_{\bb}^\phi  & = 
 \int_{q_\perp} \frac{1}{4 \pi K(q_\perp)}
\left( \sum_a \beta_a^2 + 2 \sum_{a < b} \beta_a \beta_b 
{\rm cos}(q_\perp(a-b)) \right)
\label{8}
\end{align}

\noindent where $\int_{q_\perp} = \int_{-\pi}^\pi \frac{d q_\perp}{2 \pi}$.
Similary,  a vertex operator in the dual variables, i.e.  
$V_{\bb}^\theta = {\rm exp}(i \sum_a \beta_a \theta_a)$, 
has a scaling dimension  $\Delta_{\bb}^\theta$ given 
by the formula (\ref{8}) with $K(q_\perp)$ replaced  by its inverse.
The interaction Lagrangian  
is given by the pair hopping (SC) and particle-hole (CDW) operators 
\cite{emery00,carpentier},

\begin{align}
{\cal L}_{int} = & \frac{1}{2} 
\int \frac{d^2 x}{(2 \pi a_0)^2} 
\sum_{a,b} \left[ g_{CDW}^{a,b} \; {\rm cos}\beta (\phi_a - \phi_b) 
\label{9} \right. \\
& \left. +  g_{SC}^{a,b} \;  {\rm cos}\beta (\theta_a - \theta_b) \right]
\nonumber 
\end{align}

\noindent where $a_0$ is the lattice spacing  and  
$\beta= \sqrt{2 \pi}$ for 
the charge modes of spin gapped systems \cite{emery00}, or 
$\beta = \sqrt{4 \pi}$ for spinless fermions \cite{carpentier}.
In the latter references the stability of the perturbed 
SLL was studied in terms of the relevance or irrelevance of the CDW and SC
operators (\ref{9}), given by their scaling dimensions. 
There are however cases where one has to consider the
renormalization of the functions $K(q_\perp)$ and $v(q_\perp)$,
for example when the CDW and SC operators become marginal. 
The latter situation actually arises in the study of the
coupled XXZ spin chain Hamiltonians via Ising and
spin-pair-flipping terms \cite{jose1,tpa},

\begin{align}
&  H =   \sum_{i=1}^L \sum_{a=1}^N 
\left[ - \frac{J}{2} ( S^+_{i,a} S^-_{i+1,a} + 
 S^-_{i,a} S^+_{i+1,a} )  + J_z S^z_{i,a} S^z_{i+1,a} \right. 
\label{10} \\
& \left. +  J'_z 
S^z_{i,a} S^z_{i,a+1} + J'_{XY} 
( S^+_{i,a} S^+_{i+1,a} S^-_{i,a+1} S^-_{i+1,a+1} + h.c.)
\right]
\nonumber 
\end{align}

This model can be Jordan-Wigner transformed into a spinless
fermion Hamiltonian which, after bosonization, becomes
at half filling \cite{jose1}, 

\begin{align}
& H=   \sum_{a=1}^N \int dx \left[ 
\frac{u}{2 K_0} \pi_a^2 + \frac{u K_0}{2} (\partial_x \phi_a)^2 
 + \frac{2 J_z a_0}{(2 \pi a_0)^2} {\rm cos}\sqrt{16 \pi} \phi_a
\right. \nonumber \\
&  + \frac{J'_z {a_0}}{\pi} \partial_x \phi_a \partial_x \phi_{a+1} 
+ \frac{8 J'_{XY}a_0 }{(2 \pi a_0)^2} {\rm cos} \sqrt{4 \pi} (
\theta_{a}  - \theta_{a+1}) +  \label{11} \\
& \left. 
\frac{ 2 J'_z a_0}{ (2 \pi a_0)^2} \left( 
{\rm cos} \sqrt{4 \pi} (\phi_a  + \phi_{a+1})    
- {\rm cos} \sqrt{4 \pi} (\phi_a  - \phi_{a+1}) \right) \right]   
\nonumber
\end{align} 

\noindent where $u = J a_0 (1 + 2 J_z/ (\pi J)$ and 
$K_0 = 1 + 2 J_z/(\pi J)$ for $|J_z| << J$.  
The gaussian terms in (\ref{11}) yield a SLL model 
(\ref{2}) with 

\begin{align}
&  K^J_{a,b} = \delta_{a,b} K_0  ,\;\; 
K^\rho_{a,b} = \delta_{a,b} K_0 + \frac{J'_z}{\pi} \delta_{|a- b|,1}
\label{12} 
\end{align}

\noindent  where  the time variable and the  exchange
couplings are measured in units of $u$ and $J$ respectively.
In the weak coupling regime  $|J_z| << J$, the 
SLL function $K(q_\perp)$ is close to 1. Consequently 
the intra-chain umklapp couplings ${\rm cos}\sqrt{16 \pi} \phi_a$
have  dimension $d \sim 4$ and hence can be neglected, 
as it is the case of decoupled single chains. On the
other hand the CDW, SC and Umklapp inter-chain couplings
are marginal and  one has to consider their running
together with that of the SLL functions. To simplify matters
we shall neglect in what follows the Umklapp term, which
is anyway absent away from half-filling. The couplings constants in 
(\ref{9}), which correspond to (\ref{11}),   are given by

\begin{equation}
g_{CDW}^{a,b} = - 2 J'_z \delta_{|a-b|, 1} , \;\;
g_{SC}^{a,b} = 8 J'_{XY} \delta_{|a-b|, 1} 
\label{13}
\end{equation}

Eqs.(\ref{2}) and (\ref{9})
define  a  multicomponent sine-Gordon model
which can be renormalized 
using operator product expansion (OPE)  techniques  \cite{cardy}. 
We shall further  
assume that the sound velocity of the 
modes is constant and set it to 1, 
so that the hard disk regularization will be common to all
the modes. The RG-flow equations for 
$K_{a,b} = K_{a,b}^{J,\rho}$ and $g_{CDW/SC,n}$
are given by ( up to second order in the $g's$)

\begin{align}
\frac{d K_{a,b}}{d s} &  =   \frac{\beta^2}{(4 \pi)^3} 
\left[ F_{CDW}^0 \delta_{a,b} - (g_{CDW}^{a,b})^2 
\nonumber \right. \\ 
&  \left. - \sum_{c,d} K_{a,c} ( F_{SC}^0 \delta_{c,d} - (g_{SC}^{c,d})^2 )
 K_{d,b}  \right]
\label{14} \\
\frac{d g^{a,b}_X}{d s} & =  (2 - \Delta_X^{a,b}) g_X^{a,b} 
- \frac{1}{4 \pi} F_X^{a,b} \;\;\;    (X= SC, CDW)  \nonumber
\end{align}

\noindent where $\Delta_X^{a,b}$ are the scaling dimensions
of the CDW and SC operators ( eq.(\ref{8}))

\begin{align}
& \Delta_{SC/CDW}^{a,b} = \frac{\beta^2}{2 \pi} \int_{q_\perp} 
(1 - {\rm cos} (q_\perp (a - b)) K^{\pm 1}(q_\perp) \label{15}
\end{align}

\noindent 
and  the functions $F_X^0$ and  $F^{a,b}_X$ 
are defined as,

\begin{align}
& F_X^0= \sum_b (g^{a,b}_X)^2, \;\; F_X^{a,b} = \sum_c g_X^{a,c} g_X^{c,b} 
\label{16}
\end{align}

\noindent 
As an application of these eqs. we shall consider
the case when $K(q_\perp) = 1 + k_0 + \sum_{n > 0}
k_n {\rm cos } (q_\perp n) $ is close to 1 and
$\beta = \sqrt{4 \pi}$, which corresponds
to the spinless fermion model (\ref{10}, \ref{11}).
The linearization of eqs.(\ref{14}) yield

\begin{align}
& \frac{d k_0}{d s}  =   \frac{1}{8 \pi^2} \sum_{n > 0}
( g_{CDW,n}^2 - g_{SC,n}^2 ) \nonumber \\
& \frac{d k_n}{d s}  =   - \frac{1}{8 \pi^2} 
( g_{CDW,n}^2 - g_{SC,n}^2 ), \;\; ( n > 0) \label{17} \\
& \frac{ d g_{CDW, n} }{d s }  = 
  (2 k_0 - k_n) g_{CDW,n} -  \frac{1}{4 \pi} 
\frac{ \partial {\cal N}(g_{CDW})  }{\partial 
g_{CDW, n}} \nonumber \\
& \frac{ d g_{SC, n} }{d s }  = 
 - (2 k_0 - k_n) g_{SC,n} -  \frac{1}{4 \pi} 
\frac{ \partial  {\cal N}(g_{SC}) }{\partial 
g_{SC, n}} \nonumber 
\end{align}

\noindent where ${\cal N}(g)$ is the cubic polynomial

\begin{equation}
{\cal N}(g) = \sum_{n, m >0} g_n g_m g_{n+m}
\label{18}
\end{equation}

\noindent which encodes the OPE of the CDW and SC operators appearing
in (\ref{9}). Notice the invariance of (\ref{18}) under
the change $g_n \rightarrow (-1)^n g_n$. 
From  (\ref{17}) it follows the RG conservation
of $K(q_\perp = 0) = 1 + \sum_{n \geq 0} k_n$. Moreover the
``Minkowskian'' norm $\mu$, defined as, 

\begin{equation}
\mu = k_0^2 + \frac{1}{2} \sum_{n>0} k_n^2 - \frac{1}{(4 \pi)^2 }
\sum_{n >0} ( g_{CDW,n}^2 + g_{SC,n}^2 ) 
\label{19}
\end{equation}

\noindent satisfies the RG eq.

\begin{equation}
\frac{ d \mu}{d s} = \frac{6}{(4 \pi)^2} \left[ {\cal N}(g_{CDW}) 
+ {\cal N}(g_{SC}) \right]
\label{20}
\end{equation}

Eqs.(\ref{17}) define an RG-flow in an infinite dimensional
space formed by all the SLL, CDW and SC parameters. 
To extract some information from them  
we have to do further approximations. Let us first consider
the simplest case where only $k_0$ and $g_{CDW,1}$ are non vanishing.
A truncation of eqs.(\ref{17}), written  in the  variables
$x = - 2 k_0$ and $y = g_{CDW,1}/(2 \pi)$,  gives

\begin{equation}
\frac{d x}{d s} = - y^2, \;\; \frac{d y}{d s} = - x y 
\label{21}
\end{equation}

\noindent which are the well known Kosterlitz-Thouless RG eqs. 
of the XY model \cite{KT,cardy}. 
In this case the norm $\mu$ (\ref{19})  is proportional
to $x^2 - y^2$,  and it is conserved since the cubic term
vanishes. The RG trayectories are 
hyperbolas of constant $\mu$ which end up at infinity provided
$|y| > x$. The latter inequality translates in the  XXZ+Ising model
(\ref{10}) 
into the condition $|J'_z| > - 4 J_z$, which coincides with the one
obtained in reference \cite{jose1}  
for the existence of an AF-Ising  phase. 
In  the weak coupling
regime of the XY model, which corresponds to the region 
 $|y| < x$,  the parameter $y$  flows
to zero and $x$  flows to the line $x > 0$ 
of SLL stable fixed points. The inclusion of the interstripe
Luttinger parameter
$k_1$ modifies the boundaries of the strong coupling region.
Indeed  
performing the change of variables $x = - 2 k_0 + k_1$
and $y = \sqrt{6} \;  g_{CDW,1}/(4 \pi)$ one obtains from  (\ref{17})
the same KT eqs.(\ref{21}),  and   the strong coupling regime
is given now by $J'_z > - 17.798 J_z $ ( if $J'_z > 0)$
and  $-J'_z > - 1.798 J_z $ ( if $J'_z < 0)$.
In all these cases a necessary condition to achieve 
a XY or SLL phase is to have a ferromagnetic intrachain coupling
$J_z < 0$, which  agrees with the results of \cite{carpentier}
concerning  the proximity of the stable SLL to the isotropic 
ferromagnetic point where the boson stifness $K_0$ vanishes.  
To consider the effect on the higher order couplings
we make the change of variables $x= - 2 k_0$ and
$y_n = g_{CDW,n}/(2 \pi)$, obtaining the eqs.

\begin{equation}
\frac{d x}{d s} = -\sum_{n =1}^M  y_n^2, \; \;\; 
\frac{d y_n}{d s} = - x y_n - \frac{1}{2} \frac{ \partial {\cal N}(y)}{
\partial y_n}  
\label{22} 
\end{equation}

\noindent If the initial conditions are set to
$x(0) = x, \;  y_1(0)=y, \;  y_n(0) = 0 \;\; ( 1 < n \leq M)$ we obtain
numerically that the strong coupling regime arises
when $|y_1| > A_M x$,  where $A_M$  converges
exponentially to 0.83 in the limit $M \rightarrow \infty$.
Hence the AF-Ising  region of \cite{jose1}
gets enlarged by the inclusion of all the couplings
generated by the RG, while SLL region diminishes. 
One can similarly add the effect of the $k_n$ 
parameters which change again the boundaries of the
AF-Ising region. 

Another interesting case is when 
there are both CDW and SC couplings. As can be seen seen
from eqs.(\ref{17}) the SLL fixed point, where 
$g_{CDW,n} = g_{SC,n}=0$, is unstable since their
scaling dimensions are $\pm (2 k_0 - k_n)$. However
there is another fixed point if 
$g_{CDW,n} = - g_{SC,n}$ or 
$g_{CDW,n} = (-1)^{n+1}  g_{SC,n}$ and 

\begin{equation}
(2 k_0 - k_n) g_{CDW,n} = \frac{1}{4 \pi} \frac{ \partial 
{\cal N}(g_{CDW})}{\partial g_{CDW,n}} 
\label{23}
\end{equation}

This equation has non trivial solutions for $g_n$ 
as functions of $k_n$. 
The fixed points we have analized 
turn out to be unstable. The non-gaussian nature
of these fixed points recall the
Wilson-Fisher fixed point,  with $2 k_0 - k_n$
playing the role of $\epsilon = 4-D$.
From a completely different viewpoint
it is worth to mention that the mathematical 
structure underlying  these fixed points,
is very close
to the one that appears in 
5D-Supergravity theories \cite{5d}, 
where the manifold of   the scalar fields, that
come in the Maxwell multiplets, is defined
in terms of a cubic norm  
 ${\cal N}$ similar to the one introduced
in (\ref{18}). This connection may help to study  
the properties of the
new solutions. An  interesting problem is to investigate 
the existence of stable non-gaussian SLL fixed points
when $K(q_\perp)$ does not lie near 1. This can be done 
starting from eqs.(\ref{14}) rather than from their
linearized version (\ref{17}). 
Another issue is to include
the effect of non constant $v(q_\perp)$. The 
generalization of our results to SLL's with
charge and spin degrees of freedom is rather straightforward.
Here too we expect the appearance  of novel  non-gaussian 
fixed points.

\subsection*{Acknowledgments}

I would like to thank M.A. Mart\'{\i}n-Delgado and 
J. Rodriguez for conversations. 
This work has been supported by the Spanish grant 
BFM2000-1320-C02-01.


\begin{thebibliography}{999}

\bibitem{emery} V.J. Emery, in  ``Highly Conducting 
one-dimensional solids'', J. Dovrese et al. (Plenum, New York, 1979). 


\bibitem{schulz} H.J. Schulz, J. Phys. {\bf C 16}, 6769 (1983).  


\bibitem{emery00} V.J. Emery, E. Fradkin, S.A. Kivelson
and T.C. Lubensky,  Phys. Rev. Lett. {\bf 85}, 2160 (2000). 

\bibitem{carpentier} A. Vishwanath and D. Carpentier,
Phys. Rev. Lett. {\bf 86}, 676 (2001). 

\bibitem{lubensky01} R. Mukhopadhyay, C.L. Kane, T.C. Lubensky,
cond-mat/0102163. 


\bibitem{anderson} P.W. Anderson, Phys. Rev. Lett. {\bf 67},
3844 (1991). 

\bibitem{dna} C. S. O'Hern, T. C. Lubensky, J. Toner, 
Phys. Rev. Lett. {\bf 83}, 2745 (1999).

\bibitem{KT} J.M. Kosterlitz and D.J. Thouless, J. Phys. {\bf C 6},
1181 (1973). 


\bibitem{jose1}  J.P. Rodriguez, P.D. Sacramento, V. Vieira,
Phys. Rev.  {\bf B 56}, 13685 (1997)                

\bibitem{jose2} J.P. Rodriguez,  Phys. Rev. {\bf B 58},  944 (1998). 

\bibitem{tpa}  M.A. Martin-Delgado, M. Roncaglia, G. Sierra,
cond-mat/0101458. 


\bibitem{cardy} J. Cardy, ``Scaling and Renormalization 
in Statistical Physics'', Cambridge University Press (1996).


\bibitem{5d} M. Gunaydin, G. Sierra and P. Townsend,
Nucl. Phys. {\bf B 253}, 573 (1985). 









\end{thebibliography}
\end{document}